\documentclass[manuscript]{aastex62}

\graphicspath{{./}{figures/}}

\shorttitle{Asymmetric distribution of weak photospheric magnetic field values}
\shortauthors{Getachew et al.}

\begin{document}

\title{\textbf{Asymmetric distribution of weak photospheric magnetic field values}}

\correspondingauthor{Tibebu Getachew}
\email{tibebu.ayalew@oulu.fi}

\author{Tibebu Getachew}
\affil{ReSoLVE Centre of Excellence, Space Climate Research Unit, University of Oulu, 90014 Oulu, Finland}

\author{Ilpo Virtanen}
\affiliation{ReSoLVE Centre of Excellence, Space Climate Research Unit, University of Oulu, 90014 Oulu, Finland}

\author{Kalevi Mursula}
\affiliation{ReSoLVE Centre of Excellence, Space Climate Research Unit, University of Oulu, 90014 Oulu, Finland}

\begin{abstract}
We use the synoptic maps of the photospheric magnetic field observed at Wilcox Solar Observatory, Mount Wilson Observatory, Kitt Peak, SOHO/MDI, SOLIS/VSM, and SDO/HMI to study the distribution of weak photospheric magnetic field values in 1974-2018. We fit the histogram distribution of weak field values for each synoptic map of the six data-sets separately with a parametrized Gaussian function in order to calculate the possible shift (to be called here the weak-field asymmetry) of the maximum of the Gaussian distribution from zero. We estimate the statistical significance of the weak-field asymmetry for each rotation. We also calculate several versions of lower-resolution synoptic maps from the high-resolution maps and calculate their rotational weak-field asymmetries. We find that the weak-field asymmetries increase with decreasing map resolution. A very large fraction of weak-field asymmetries are statistically significant, with the fraction of significant weak-field asymmetries increasing with decreasing resolution. Significant weak-field asymmetries of high- and low-resolution maps mainly occur at the same times and have the same sign. Weak-field asymmetries for the different data-sets and resolutions vary quite similarly in time, and their mutual correlations are very high, especially for low-resolution maps. These results give strong evidence for weak-field asymmetries reflecting a real feature of weak field values, which is best seen in medium- and low-resolution synoptic maps and is most likely related to the supergranulation scale of the photospheric field.
\end{abstract}

\section{Introduction}
\label{introduction}

The Sun's photospheric magnetic field is the source of the coronal and heliospheric magnetic fields \citep[see, e.g.,][]{Wilcox1968,Svalgaard1976,Svalgaard2011,Getachew2017}. Detailed study of the fundamental properties of the solar photospheric magnetic field is crucial to understand the Sun's radiative and particle outputs that affect the Earth's near-space environment, as well as the entire heliosphere. Photospheric magnetic field data is an essential parameter for space weather and space climate. Solar magnetic field models are used to extrapolate the photospheric magnetic field into the corona and heliosphere.\\
 
The large-scale photospheric magnetic field has been measured since 1950s \citep{Babcock1953}. However, routine photospheric magnetic field observations began at the Mount Wilson Observatory (MWO) in the 1970s \citep{Howard1983,Howard1989}, which stopped  observations in January 2013. Observations of the photospheric magnetic field have also been made at the National Solar Observatory \citep[NSO,][]{Livingston1976} and at the Wilcox Solar Observatory (WSO) since the 1970s \citep{Svalgaard1978,Hoeksema2010}. At NSO Kitt Peak Vacuum Telescope (KPVT) two instruments were used: the 512-channel magnetograph in 1975{\thinspace}--{\thinspace}1993 and spectromagnetograph in 1992{\thinspace}--{\thinspace}2003. KPVT was succeed by Synoptic Optical Long-term Investigations of the Sun (SOLIS) telescope and the Vector Spectromagnetograph (VSM) in 2003. Space-based high resolution magnetic field observations were obtained from the Michelson Doppler Imager (MDI) instrument on-board the Solar and Heliospheric Observatory (SOHO) spacecraft from 1996{\thinspace}--{\thinspace}2011 \citep{Scherrer1995}. Space-based high resolution vector magnetic field observations are provided by the Helioseismic and Magnetic Imager (HMI) on-board the Solar Dynamics Observatory (SDO) spacecraft since 2010 \citep{Schou2012,Hoeksema2014}. \\

The existing photospheric magnetic field data-sets are produced using different instrumentations and mostly using different measurement techniques. Therefore, the magnetic field intensity varies significantly between the observatories. Several studies have been carried out to compare the different photospheric magnetic field data-sets and their mutual relationship \citep[see, e.g.,][]{Riley2014,Virtanen2017}.\\

In this article we study the distribution of weak photospheric magnetic field values using several different data-sets. 
The maximum of the weak-field distribution is often slightly shifted from zero. The non-zero peak location, also called the zero-level offset, is commonly considered to result from the many observational challenges and problems related to the magnetograph instruments. 
Based on an earlier idea by \citet{Ulrich2002}, \citet{Liu2004} used a two-fold method to remove the zero-level offset from the magnetic field observations. They first made a Gaussian fit to the weak field values in order to find the offset of each image and then, by high-pass filtering, removed the highly-fluctuating (random) part of the zero-level offset values. This method has been used, at least, to correct the magnetic field observations made by the SOHO/MDI and SDO/HMI instruments. All magnetograph instruments correct their observations for the zero-level offset, although the related methods are not always very well documented. Moreover, at all magnetograph observatories, the synoptic maps are constructed from the zero-level offset corrected images. \\

In analogy with the method of \citet{Liu2004}, we fit a Gaussian function to the histogram distribution of weak field values in order to study the possible shift of weak field values of synoptic maps. If the maximum of the fitted distribution is not at zero field value, the corresponding non-zero field value of the distribution maximum is defined as the weak-field asymmetry. 
Despite the similar method, we use a different term for the offset (shift) in order not to confuse the current shifts with the (already removed) zero-level offsets. We apply this method for the synoptic maps of different data-sets and make a comparative study of the obtained weak-field asymmetries. We also study weak-field asymmetry values for different resolutions of a given data-set by changing the resolution of the synoptic maps. Our results show that the weak-field asymmetries reflect a real feature of the distribution of weak photospheric magnetic field values. 
The paper is organized as follows. Section ~\ref{Data and methods} presents the data and methods used. Section ~\ref{weak-field asymmetries of HMI synoptic maps} focuses on the weak-field asymmetries of HMI data and Section ~\ref{weak-field asymmetries of MDI synoptic maps} discusses the weak-field asymmetries of MDI data. Section ~\ref{weak-field asymmetries of KPVT and VSM synoptic maps} discusses the weak-field asymmetries of KPVT and SOLIS/VSM data. Sections ~\ref{weak-field asymmetries of MWO synoptic maps} and ~\ref{weak-field asymmetries of WSO synoptic maps} present the weak-field asymmetries of MWO and WSO data, respectively. We compare the weak-field asymmetries of the different data-sets in Section ~\ref{Comparing the weak-field asymmetry values of the six data-sets}. Finally, we discuss the results and give our conclusions in Section ~\ref{Discussion and conclusions}.\\

\section{Data and methods}
\label{Data and methods}

In this paper we use measurements of the photospheric magnetic field at WSO, MWO, KPVT, MDI, SOLIS/VSM, and HMI to calculate the weak-field asymmetry. (See later for a brief review of each data-set. Detailed reviews of these data-sets can be found, e.g., in \citet{Riley2014} and \citet{Virtanen2016}). We fit the histogram distribution of measured field values for each synoptic map of the six data-sets separately with a parametrized Gaussian function to calculate the position of the peak of the Gaussian distribution. The parametrized Gaussian function given by

\begin{equation}
\phi =c\exp\left({-\frac{1}{2}\left(\frac{B_{i}-a}{b}\right)^{2}}\right)
\label{equ:Gaussian}
\end{equation}
 is fitted to the histogram distribution, where $B_{i}$ are the field values, and the three parameters $a$, $b$ and $c$ give the position of the peak (the weak-field asymmetry), the width of the Gaussian distribution (standard deviation) and the amplitude, respectively. The three model parameters are determined using least squares regression based on Gauss-Newton method.\\

 We calculated the fitted values of $a$ (hereafter called weak-field asymmetry values) for all synoptic maps of all six data-sets, and studied their long-term evolution over several solar cycles. Note that synoptic maps are constructed from daily magnetograms that, before construction, have been corrected for the (instrumental) zero-level offsets. In order to assess the significance of the weak-field asymmetry, we applied Student's t-distribution given as
 \begin{equation}
\hat{t}=\frac{{\mu}}{S_{\mu}}
\label{equ:t_test}
\end{equation}
where $S_{\mu}$ is the error of the mean $\mu$ (the fit value of $a$). We compared the calculated $\hat{t}$ value to the corresponding precalculated statistic, $t_{\alpha,n-3}$ value, where $\alpha$ is the significance level and $n= 6001$ (for HMI high resolution data). The value of $\mu$ is statistically significant at $99\%$ confidence level, if $\hat{t}>t_{0.01,n-3}$ (for a $99\%$ confidence interval, $\alpha=0.01$). \\

To compare weak-field asymmetry values between low- and high-resolution versions of a given synoptic map, block averaging method is applied. Let the (original) high resolution map consist of $N=N_{\phi}*N_{\theta}$ pixels, where $N_{\phi}$ is the number of longitude pixels and $N_{\theta}$ the number of sine-latitude pixels. Then the high resolution map is averaged to blocks of $L$ pixels, where $L<N$ to give a map that has $N/L$ pixels, where the pixels of the low-resolution map have a larger area than the pixels of the high-resolution map. It is worth noting that converting the high-resolution map to a low-resolution synoptic map by averaging magnetic field values to larger pixels does not give exactly the same result as original observations at lower resolution. In other words, block-averaging the magnetic field of the high-resolution synoptic data over a particular area does not yield exactly the same field strength as obtained from the line shift over that area \citep[see also][]{Riley2014}. In addition, block-averaging may sum up some of the large field values of the high-resolution maps to (some of the) small field values of the low-resolution maps, which modifies the distribution of the weak-field values.\\

\section{Weak-field asymmetries of HMI synoptic maps}
\label{weak-field asymmetries of HMI synoptic maps}

 In this paper we use the 3600*1440 HMI radial-field synoptic maps (equally spaced in longitude and sine-latitude, whence each pixel of the synoptic map represents the same surface area on the solar surface). In the HMI data pipeline, the 3600*1440 synoptic maps are constructed as follows \citep{Liu2012,Hoeksema2014,Hayashi2015}: The 4096*4096 HMI 720s LOS magnetograms are first converted to pseudo-radial field magnetograms by assuming that the photospheric field is approximately radial. Pseudo-radial magnetograms are then remapped and interpolated onto a very high-resolution Carrington coordinate grid. The possible zero-level errors of these high-resolution HMI magnetograms are removed using the method proposed by \cite{Liu2012}, which was first applied to MDI data (see later in Sections Section ~\ref{weak-field asymmetries of MDI synoptic maps}). Zero-offset corrected pseudo-radial field magnetograms in the Carrington system are then averaged to give the 3600*1440 pseudo-radial synoptic map. The field values at each longitude of the radial synoptic map are the average of the field values nearest to the central meridian from the 20 best 720s cadence magnetograms, covering 4 hours (equivalent to $2.2 ^{o}$ longitude), i.e., within $\pm1.1^{o}$ of central meridian. The HMI synoptic maps included in this paper cover Carrington rotations (CR) $2097-2207$, i.e., the time interval from 2010.4{\thinspace}--{\thinspace}2018.7. The synoptic maps used in this paper include only the visible areas of the solar disk, i.e, poles are not filled at times when the polar regions are partly invisible due to  the $\pm7.25 ^{o}$ tilt of the Earth’s orbit with respect to the heliographic equatorial plane.\\

\begin{figure}[htpb] 
	\centering
	\includegraphics[width=\columnwidth]{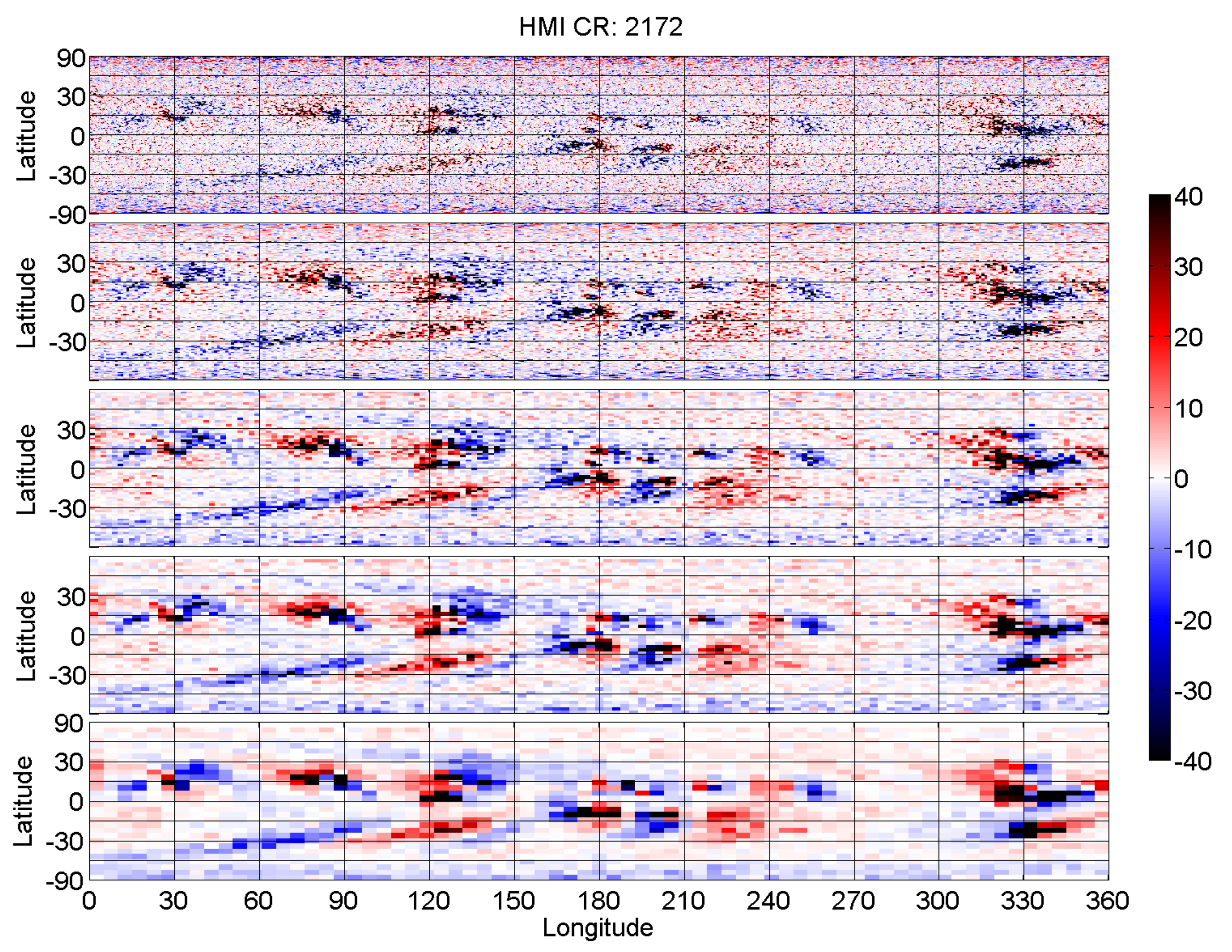}
	\caption{HMI synoptic map for CR 2172 at different resolutions. From top to bottom: 3600*1440, 360*180, 180*72, 120*48 and 72*30 resolution maps.}
	\label{fig:hmi_cr_2172}
\end{figure}

We derived different sets of medium- and low-resolution HMI synoptic maps (360*180, 180*72, 120*48, and 72*30) from the original HMI 3600*1440 synoptic maps to investigate the effect of data resolution on the weak-field asymmetries. The values of each pixel of the 360*180, 180*72, 120*48, and 72*30 resolution synoptic maps are block-averages of the values of $10*8$, $20*20$, $30*30$ and $50*48$ pixels in the original synoptic map, respectively. Figure \ref{fig:hmi_cr_2172} shows an example of an HMI synoptic map at the five different resolutions for CR 2172 (December 25, 2015 {\thinspace}--{\thinspace} January 21, 2016). As can be seen in Figure \ref{fig:hmi_cr_2172}, the five synoptic maps are quite similar, but the fine structure of the field is more visible as the resolution increases, as expected. On the other hand, the large-scale structure of the field, in particular the spatial extent and magnetic polarity of the largest active regions and the polar regions, are more clear in the low-resolution maps.\\

\begin{figure}[htpb] 
	\centering
	\includegraphics[width=\columnwidth]{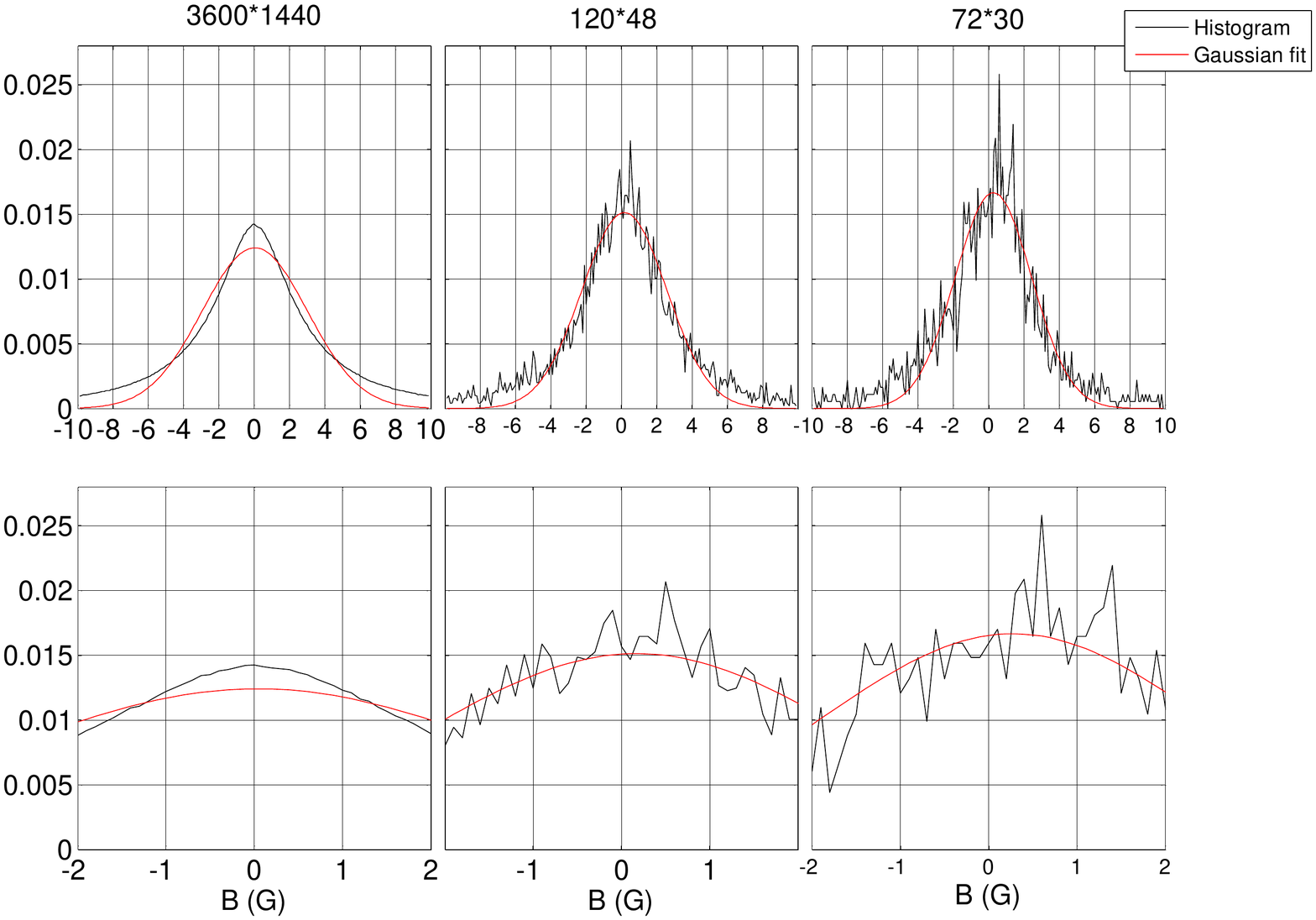}
	\caption{Distribution of field values of HMI synoptic maps for CR $2172$ at three different resolutions. Black curve shows the histogram distribution, and red curve the Gaussian fit. Upper row from left to right shows the distribution for 3600*1440, 120*48 and 72*30 resolution maps between $-10G$ to $+10G$. Bottom row is the blow-up of the upper row plots, for smaller field values.}
	\label{fig:hmi_hist_dist_2172}
\end{figure}

Figure \ref{fig:hmi_hist_dist_2172} shows an example of the distribution of weak photospheric magnetic field values for CR 2172 obtained from the synoptic maps of three different resolutions (3600*1440, 120*48 and 72*30). The photospheric magnetic field distribution (histogram) with a bin size of $0.1 G$ is derived for pixels that have values within $\pm 10 G$ (upper row of Figure \ref{fig:hmi_hist_dist_2172}). The bottom row of Figure \ref{fig:hmi_hist_dist_2172} depicts the blow-up of the upper row plots showing field values within $\pm 2 G$. 
(Weak-field asymmetries do not depend on the range of the fit, at least between about 5G and a few hundred G, as also earlier noted by \citet{Liu2004}). Note that, when reducing the spatial resolution of the synoptic maps, the relative fraction of small field values increases. \textbf{At the same time, the distribution becomes more Gaussian, since averaging modifies the distribution of field values towards normal distribution. This supports our usage of the Gaussian to fit the field values.} As can be seen in Figure \ref{fig:hmi_hist_dist_2172}, the maximum of the distribution for the 3600*1440 resolution synoptic map is obtained quite close to the zero-field value. Accordingly, there is no visible weak-field asymmetry in this case. The weak-field asymmetry of the distribution for the 120*48 synoptic map is about $+0.2 G$ and clearly visible in  Figure \ref{fig:hmi_hist_dist_2172} (especially lower panel). For the 72*30 resolution synoptic map, the maximum of the Gaussian distribution is even larger, about $ +0.3 G$. Note also that the maximum of the histogram distribution happens to be at about $ +0.6 G$.\\

\begin{figure}[htpb]
	\centering
	\includegraphics[width=\columnwidth]{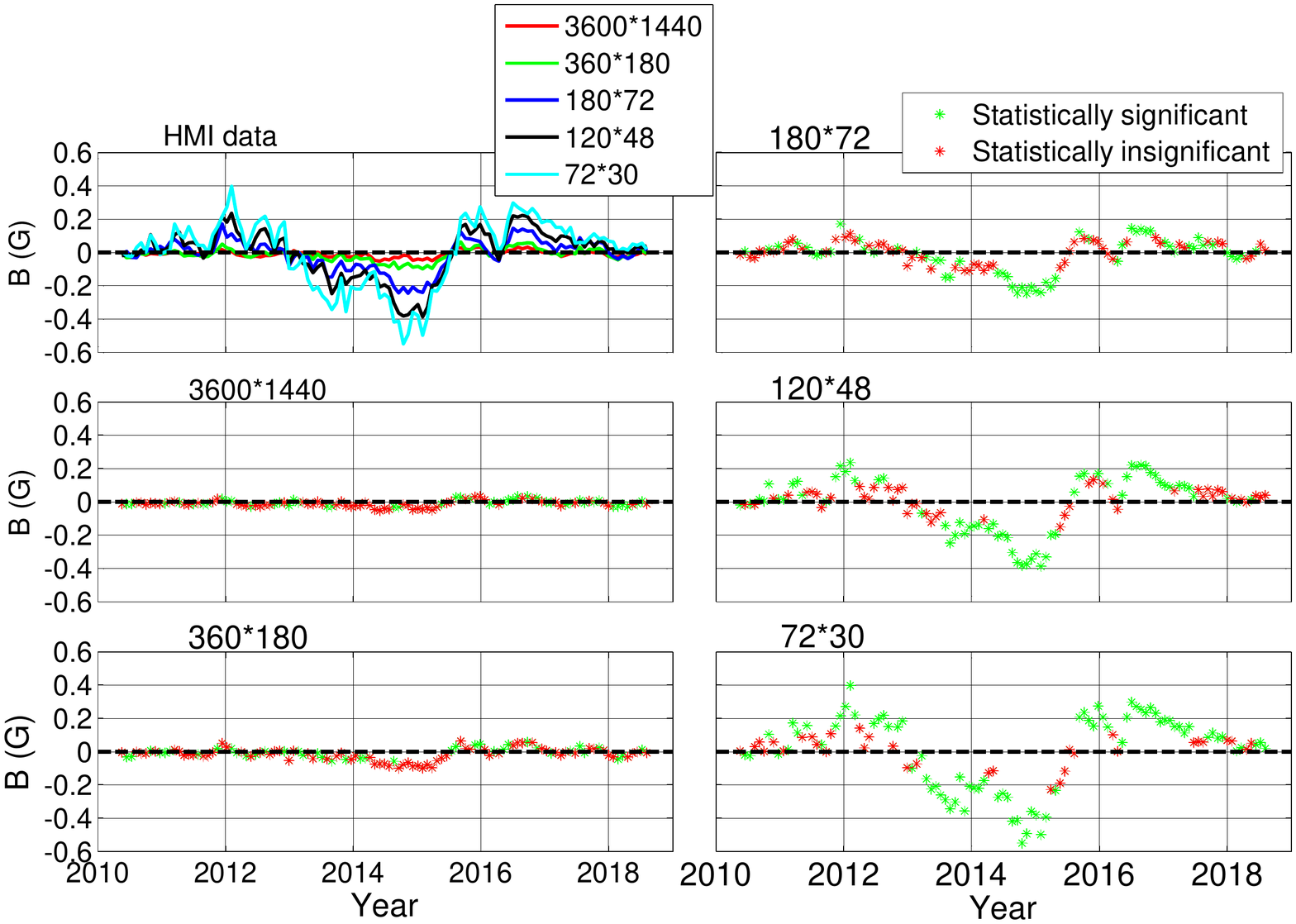}
	\caption{HMI synoptic map weak-field asymmetries for 3600*1440 (red line), 360*180 (green line), 180*72 (blue line), 120*48 (black line) and 72*30 (cyan line) resolutions are shown in the top left panel. In other five panels the statistically significant and insignificant weak-field asymmetries are noted by green and red asterisks, respectively, and are repeated in separate panels for each resolution.}
	\label{fig:hmi_zero_level_error}
\end{figure}

 Figure \ref{fig:hmi_zero_level_error} shows the HMI weak-field asymmetry values (from Gaussian fit) for each synoptic map, as well as their statistical significance at $99\%$ confidence level. The top left panel of Figure \ref{fig:hmi_zero_level_error} shows the rotational values of the weak-field asymmetries of HMI synoptic maps for all five different resolutions. Other panels of Figure \ref{fig:hmi_zero_level_error} reproduce the weak-field asymmetries from the upper left panel in order to show their statistical significance. Weak-field asymmetries which are statistically significant (significantly different from zero) are marked with green asterisk. All other weak-field asymmetries are not significantly different from zero and are marked with red asterisk. Out of 111 cases (rotations), there are 45, 46, 72, 83 and 96 statistically significant weak-field asymmetry values for 3600*1440, 360*180, 180*72, 120*48 and 72*30 resolution synoptic maps, respectively. This implies that more than $60\%$ of the weak-field asymmetry values of the 180*72, 120*48 and 72*30 resolution synoptic maps are statistically significant. Even for the two highest-resolution maps $41\%$ and $42\%$  of the weak-field asymmetry values are statistically significant. \\

As can be seen in the upper left panel of Figure \ref{fig:hmi_zero_level_error}, the values of the weak-field asymmetries are quite different for the distributions of the high- and low-resolution data. For highest-resolution HMI synoptic map (1440*3600), the weak-field asymmetries are quite small, although different from zero in $41\%$ of cases. As noted above, since the original (3600*1440) HMI synoptic maps are constructed from zero-level offset corrected magnetograms, the weak-field asymmetry of the high-resolution synoptic map is expected to remain quite small. As the spatial resolution of the data is decreased, the maximum of the field distribution shows larger weak-field asymmetry values.\\

 We note that the significant weak-field asymmetries tend to have the same sign and to occur at the same times at all resolutions despite their different absolute values. For instance, $59\%$ of significant weak-field asymmetries of the 180*72 maps occur at the same time as significant weak-field asymmetries of the 120*48 synoptic maps and $80\%$ of significant weak-field asymmetries of 120*48 maps are included within the times of the significant weak-field asymmetries of the 72*30 resolution maps. Also $70\%$ of significant weak-field asymmetries of the highest-resolution (3600*1440) synoptic maps occur at the same time as significant weak-field asymmetries of the lowest resolution (72*30) synoptic maps. This implies that weak-field asymmetries, even for the very small values of the high-resolution maps are not randomly distributed, but are mainly clustered to certain common times. As will be shown later, the weak-field asymmetries of the HMI synoptic maps also show a similar pattern with the weak-field asymmetries derived from the other instruments, which further verifies the non-random nature of the observed asymmetries.\\

The most essential feature of HMI weak-field asymmetries is the long sequence of negative values in 2{\thinspace}--{\thinspace}3 successive years in 2013{\thinspace}--{\thinspace}2015, which maximizes at the turn of 2014/2015. There is also a sequence of (slightly smaller) positive values in 2011{\thinspace}--{\thinspace}2012 and in 2015{\thinspace}--{\thinspace}2017. The length and height of the weak-field asymmetries at these times depend on the resolution of the map, reflecting the occurrence and correlation of the significant asymmetry times described above.\\

\section{Weak-field asymmetries of MDI synoptic maps}
\label{weak-field asymmetries of MDI synoptic maps}

The original MDI synoptic maps used in this paper have 3600*1080 pixels (equally spaced in longitude and sine-latitude). These synoptic maps give the pseudo-radial fields, where missing values in the polar regions are not filled. \citep[Details about MDI data can be found, e.g., in][]{Scherrer1995,Liu2004}. We briefly discuss below the construction process of MDI synoptic maps. At MDI, a one-minute magnetogram (1024*1024 resolution) is created from four filtergrams at 96 minutes cadence. Magnetograms are remapped to a high resolution Carrington coordinate grid and converted to a pseudo-radial field.\\

 \citet{Liu2004} derived the MDI weak-field shifts using the same method as used in this paper. They showed that some weak-field shifts are systematically non-zero, implying that a fraction of shifts are physically relevant. They used a high-pass filter to the time series of calculated weak-field shifts to exclude random shifts (zero-offsets). Accordingly, level 1.8 MDI magnetograms are corrected for zero-offsets, and the 3600*1080 resolution synoptic maps are derived from the zero-offset corrected pseudo-radial level 1.8 magnetograms. One pixel in MDI synoptic map is the average of nearly central meridian measurements of approximately 20 one-minute individual magnetograms, with an effective temporal width of about one day, i.e., within $\pm7 ^{o}$ of central meridian. \\

\begin{figure}[htpb]
	\centering
	\includegraphics[width=\columnwidth]{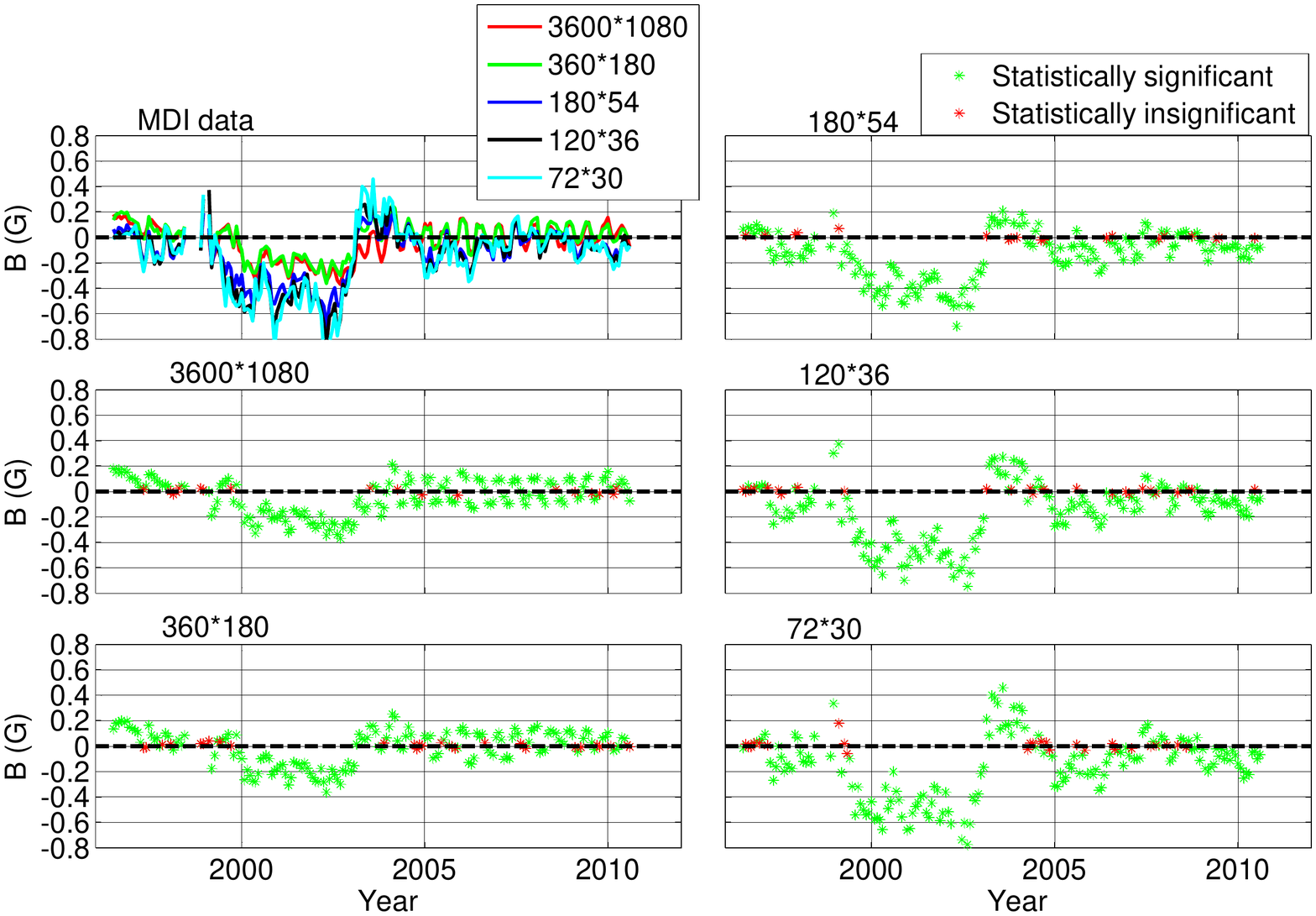}
	\caption{MDI synoptic map weak-field asymmetries for 3600*1080 (red line), 360*180 (green line), 180*54 (blue line), 120*36 (black line) and 72*30 (cyan line) resolutions are shown in the top left panel. In other five panels the statistically significant and insignificant weak-field asymmetries are noted by green and red asterisks, respectively, and are repeated in separate panels for each resolution.}
	\label{fig:mdi_zero_level_error}
\end{figure}

We calculated four sets of low-resolution MDI synoptic maps (360*180, 180*54, 120*36 and 72*30) from 3600*1080 resolution synoptic map. The MDI synoptic maps included in this paper cover CR 1909{\thinspace}--{\thinspace}2100, i.e., between 1996.3{\thinspace}--{\thinspace}2010.6. Figure \ref{fig:mdi_zero_level_error} shows the MDI weak-field asymmetries and their statistical significance at $99\%$ confidence level. The upper left panel shows the time evolution of weak-field asymmetry values derived from the five different resolution synoptic maps. Other panels of Figure \ref{fig:mdi_zero_level_error} reproduce the weak-field asymmetries from the upper left panel with their statistical significance, as for HMI in Figure \ref{fig:hmi_zero_level_error}. \\

 As can be seen in Figure \ref{fig:mdi_zero_level_error} the MDI weak-field asymmetries, especially those calculated from 3600*1080 (original) and 360*180  synoptic maps show strong semi-annual oscillation after 2003. The time intervals when the MDI weak-field asymmetries show strong semi-annual oscillation coincides with the time interval when the SOHO probe has been repeatedly rolled upside down every three months after an antenna failure. Because one (north-east) of the four quadrants of MDI is more noisy than others, the 3-monthly re-orientation of SOHO produces the observed semi-annual oscillation. Although this semi-annual oscillation has a less dramatic effect for the asymmetries of the low-resolution synoptic maps, we must consider SOHO/MDI weak-field asymmetries as largely unreliable after 2003. Moreover, problems in MDI shutter cause increased zero-level error since early 2000. It is unclear how well this error has been removed from level 1.8 magnetograms and synoptic maps. \\

 As can be seen in Figure \ref{fig:mdi_zero_level_error}, out of 82 cases (rotations) within the times between 1996.3{\thinspace}--{\thinspace}2003, more than $80\%$ of the weak-field asymmetry values of all the five synoptic maps are statistically significant and more than $72\%$ of these sets of significant asymmetries of all the five synoptic maps tend to occur at the same times. Note that all the MDI synoptic maps at five different resolutions depict negative weak-field asymmetries for 2{\thinspace}--{\thinspace}3 consecutive years in 1999{\thinspace}--{\thinspace}2002, and slightly positive weak-field asymmetry values in 2003.
 
\section{Weak-field asymmetries of KPVT and VSM synoptic maps}
\label{weak-field asymmetries of KPVT and VSM synoptic maps}

Synoptic maps of the photospheric magnetic field have been produced at the NSO Kitt Peak (KP) using KPVT telescope from CR 1625{\thinspace}--{\thinspace}2007, i.e., between 1975.1{\thinspace}--{\thinspace}2003.7 and SOLIS/VSM instrument from CR 2007{\thinspace}--{\thinspace}2196, i.e., between 2003.7{\thinspace}--{\thinspace}2017.8. KPVT telescope included the 512-channel magnetograph in 1975{\thinspace}--{\thinspace}1993 and spectromagnetograph in 1992{\thinspace}--{\thinspace}2003.7. The KPVT synoptic maps have 360*180 pixels (equally spaced in longitude and sine-latitude). We also derived 72*30 resolution synoptic maps from these 360*180 KPVT synoptic maps. Note that the KPVT synoptic maps give the pseudo-radial field and the missing values of the polar regions are filled. 

SOLIS/VSM synoptic maps are constructed from magnetograms measured under good weather conditions during a temporal window of 40 days, and several (up to 60) magnetograms are used to construct one synoptic map. The VSM synoptic map is the average of these full disk magnetograms weighted by $\cos^{4}\left(CMD\right)$, where $CMD$ is the central meridian distance. The VSM synoptic maps used in this study have 1800*900 pixels (equally spaced in longitude and sine-latitude) from which we calculated two sets of low-resolution synoptic maps (360*180 and 72*30). VSM synoptic maps give the pseudo-radial fields without polar filling. \citep[Details about VSM data can be found, e.g., in][]{Bertello2014}.  \\

\begin{figure}[htpb]
	\centering
	\includegraphics[width=\columnwidth]{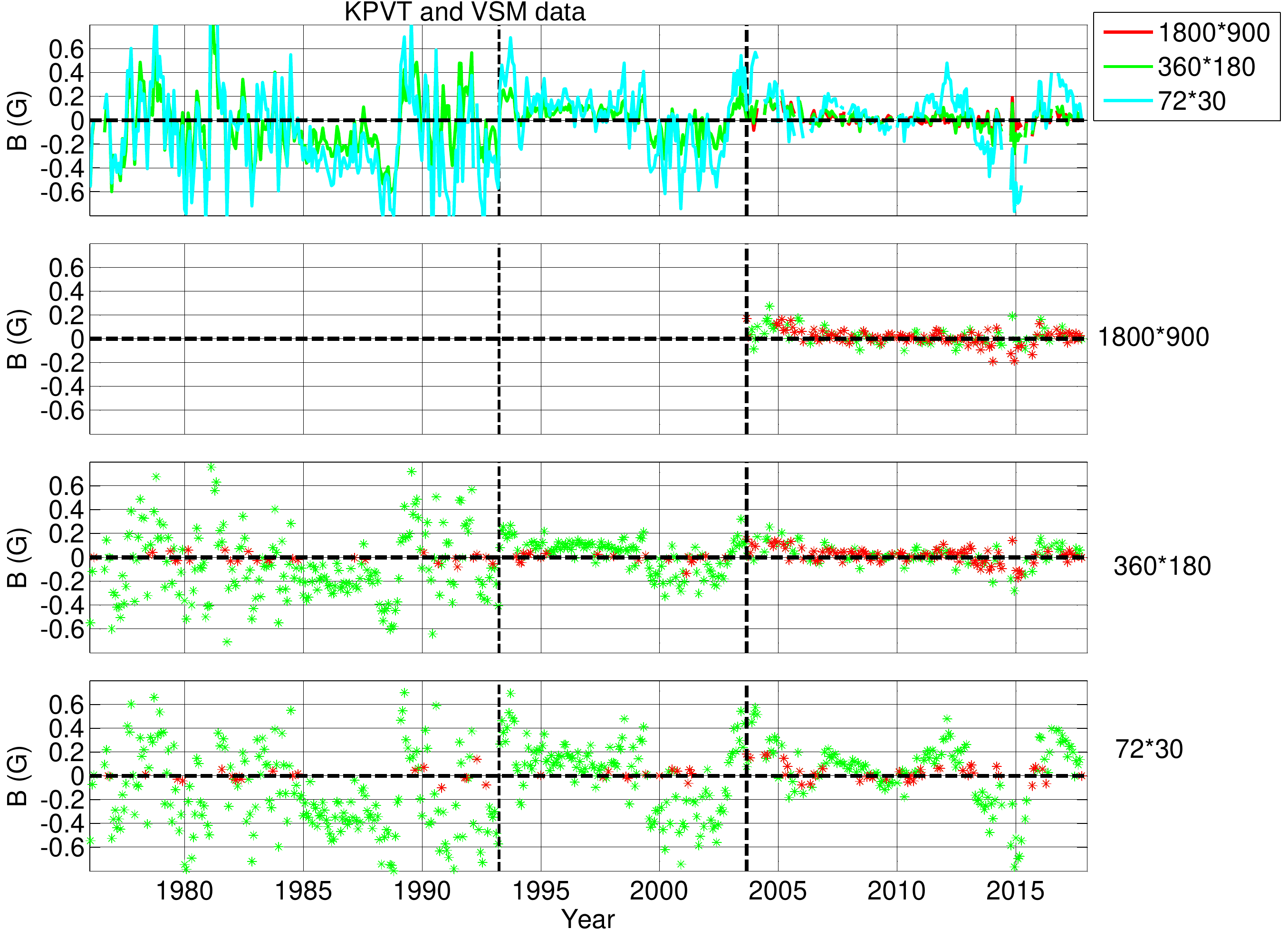}
	\caption{KPVT and VSM synoptic maps weak-field asymmetries for 1800*900 (red line), 360*180 (green line), and 72*30 (cyan line) resolutions are shown in the top  panel. In other three panels the statistically significant and insignificant weak-field asymmetries are noted by green and red asterisks, respectively, and are repeated in separate panels for each resolution. KPVT instrument update in 1993.2 is indicated by a thin vertical dashed line,  KPVT and VSM are separated by a thick vertical dashed line. }
	\label{fig:solis_zero_level_error}
\end{figure}

Figure \ref{fig:solis_zero_level_error} shows KPVT (1975.1{\thinspace}--{\thinspace}2003.7) and VSM (2003.7{\thinspace}--{\thinspace}2017.8) weak-field asymmetries and their statistical significance at $99\%$ confidence level. As can be seen from Figure \ref{fig:solis_zero_level_error}, the weak-field asymmetries of the 360*180 and 72*30 KPVT synoptic maps are closely similar except that 72*30 KPVT synoptic maps have somewhat larger values than the 360*180 maps. The weak-field asymmetry values calculated from the VSM synoptic maps greatly depend on the resolution of the data. Weak-field asymmetries are quite small in the high-resolution data (1800*900) but the values increase as the resolution of the data becomes lower. Overall, KPVT asymmetries are considerably larger than for VSM (or HMI and MDI) at similar resolution. This difference is particularly clear for 360*180 resolution. Also, the KPVT asymmetries are clearly larger during the 512-channel magnetograph period than later during the spectromagnetograph. The exceedingly large KPVT weak-field asymmetry values may be due to the well-known problems in the data of these instruments that are being corrected \citep{Harvey2015}.\\

 Out of 169 cases (rotations) in VSM maps, there are 67 ($40\%$), 74 ($44\%$) and 129 ($76\%$) statistically significant weak-field asymmetry values for 3600*1080, 360*180 and 72*30 maps, respectively. Out of $67$ significant weak-field asymmetries of the 3600*1080 maps, 33 ($49\%$) occur at the same time as significant weak-field asymmetries of the 360*180 VSM synoptic maps and out of 74 significant weak-field asymmetries of 360*180 VSM maps 55 ($74.3\%$)  are included within the times of the significant weak-field asymmetries of the 72*30 VSM maps. Interestingly, out of 67 significant weak-field asymmetries of the highest resolution (3600*1080) synoptic maps $55\%$ occur at the same time as significant weak-field asymmetries of the lowest resolution (72*30) VSM synoptic maps. Similarly, out of 377 cases (rotations) in KPVT maps, there are 328 ($87\%$) and 344 ($91\%$) statistically significant weak-field asymmetry values for 360*180 and 72*30  resolution synoptic maps, respectively. Out of $328$ significant weak-field asymmetries of the 360*180 KPVT maps, 306 ($93.3\%$) occur at the same time as significant weak-field asymmetries of the 72*30 KPVT synoptic maps.\\

Note that VSM (see Figure \ref{fig:solis_zero_level_error}) and HMI (see Figure \ref{fig:hmi_zero_level_error}) depict negative weak-field asymmetry values in 2013{\thinspace}--{\thinspace}2015 and positive values before and after this period very similarly. This is particularly clearly visible for the 72*30 resolution maps. Note that not only the timing but also the absolute values of asymmetries are very similar for this two instruments. These similarities further prove that most weak-field asymmetries are physical, not random.

\section{Weak-field asymmetries of MWO synoptic maps}
\label{weak-field asymmetries of MWO synoptic maps}

The original MWO synoptic map gives the line-of-sight photospheric magnetic field with a resolution of 971*512 evenly spaced in both longitude and latitude (rather than sine-latitude), from which we calculated the pseudo-radial field. The MWO synoptic map is constructed from 
 512*512 magnetograms corrected for instrumental zero-offsets. Magnetograms are constructed by over-sampling the original observations obtained using the 12 and 20 arcsec squared apertures. \citep[Details about MWO data can be found, e.g., in][]{Ulrich2002,Ulrich2013}. We use all the available MWO synoptic maps from 1974.5{\thinspace}--{\thinspace}2013 in their original format, where the missing values of the polar regions are not filled. We also derived other two sets of low-resolution synoptic maps (360*180 and 72*30) by first re-sampling the original synoptic data using nearest neighbor method to very high resolution (20160*10080 equally spaced longitude and sine-latitude points) and then block averaging this very high-resolution map to 360*180 and 72*30 maps. \\

\begin{figure}[htpb]
	\centering
	\includegraphics[width=\columnwidth]{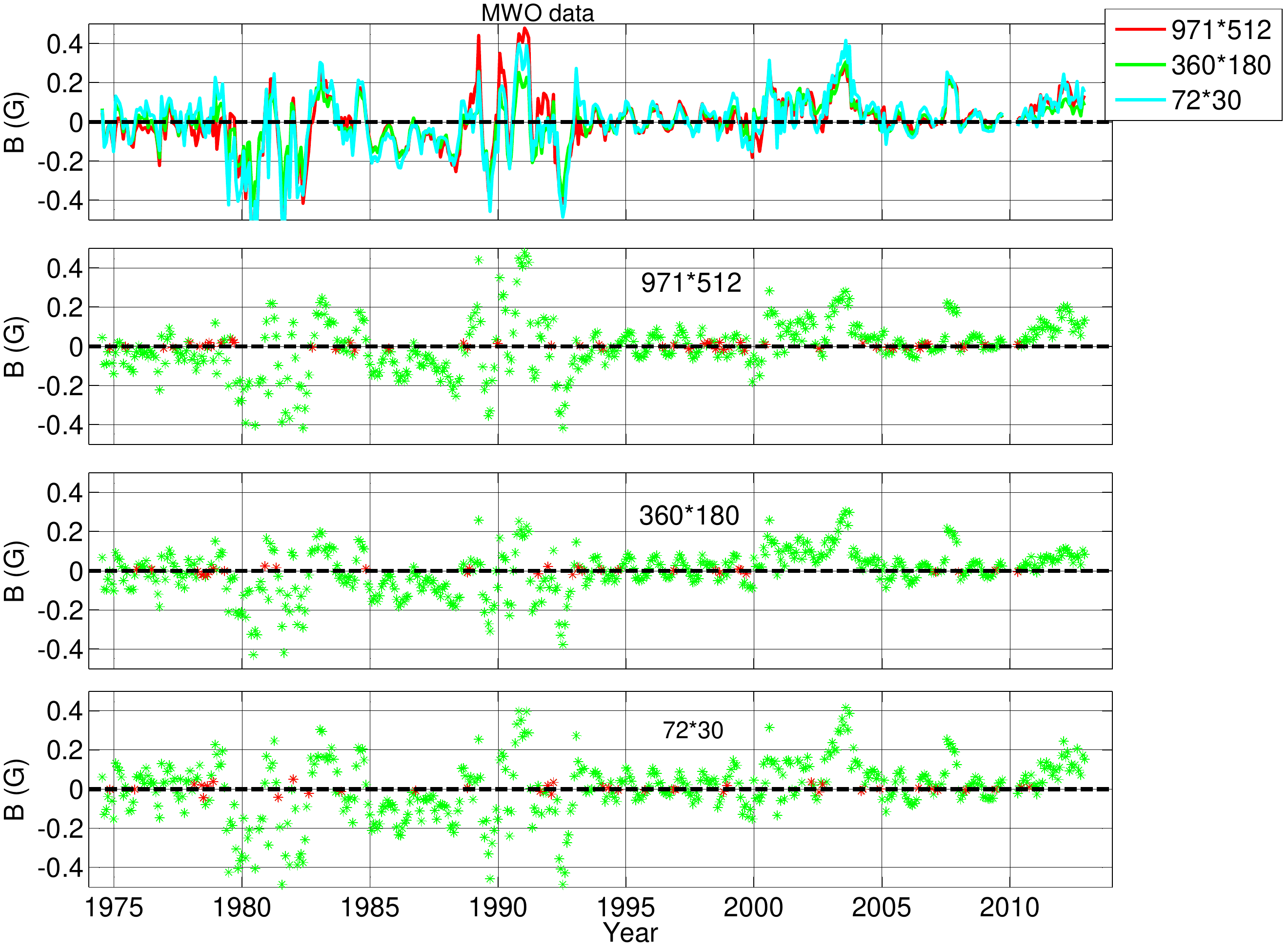}
	\caption{MWO synoptic map weak-field asymmetries for 971*512 (red line), 360*180 (green line), and 72*30 (cyan line) resolutions are shown in the top  panel. In other three panels the statistically significant and insignificant weak-field asymmetries are noted by green and red asterisks, respectively, and are repeated in separate panels for each resolution.}
	\label{fig:MWO_zero_level_error}
\end{figure}

Figure \ref{fig:MWO_zero_level_error} shows the weak-field asymmetry values of the MWO synoptic maps at three different resolutions. The weak-field asymmetries derived from the three sets of synoptic maps are surprisingly similar, the weak-field asymmetry values of the 72*30 maps being only slightly larger than the two other sets. There is no similar clear increase of asymmetries with decreasing resolution for MWO, as found especially for HMI and SOLIS/VSM. This is probably related to the lower effective resolution of MWO than HMI and VSM, with MWO high-resolution results reflecting the over-sampling. This is seen also in the very good correlation of asymmetry times. Out of 507 cases (rotations), more than $90\%$ of the weak-field asymmetry values of all the three synoptic maps are statistically significant and more than $93\%$ of these sets of significant asymmetries of all the three synoptic maps tend to occur at the same times.

\section{Weak-field asymmetries of WSO synoptic maps}
\label{weak-field asymmetries of WSO synoptic maps}

\begin{figure}[htpb]
	\centering
	\includegraphics[width=\columnwidth]{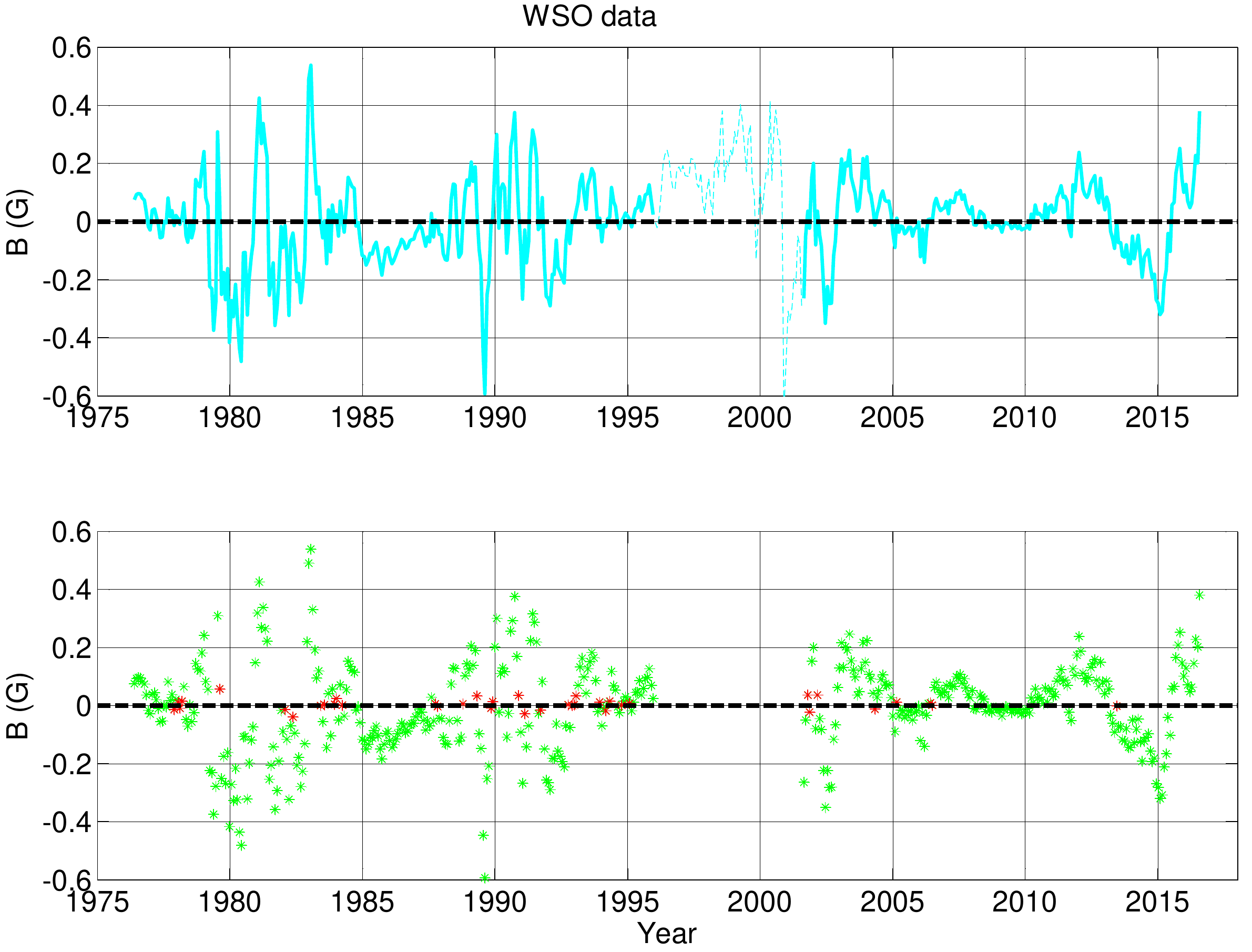}
	\caption{Weak-field asymmetries of the WSO synoptic maps. Upper panel shows rotational values of the weak-field asymmetries. Weak-field asymmetries during the erroneous data period are shown by a dashed line. Bottom panel reproduces the rotational weak-field asymmetries to show statistically significant (green asterisk) and insignificant (red asterisk) weak-field asymmetry values.}
	\label{fig:WSO_zero_level_error}
\end{figure}
  
The WSO synoptic map gives the line-of-sight magnetic field with a resolution of $72*30$ (evenly spaced in longitude and sine latitude, highest latitude bin centered at
 $\pm 75.2^{o}$). We calculated the pseudo-radial field from the line-of-sight magnetic fields. Details about the WSO synoptic maps can be found, e.g., in \citet{Hoeksema1984}. The WSO measurements are good in that the same instrumentation has been operational since measurements have started in 1976. In this paper we used all the available WSO data from CR 1642{\thinspace}--{\thinspace}2180, i.e., between 1976.3{\thinspace}--{\thinspace}2018, which makes it the only data-set to provide data continuously with no major instrument changes during the last four solar cycles.\\

Figure \ref{fig:WSO_zero_level_error} shows the weak-field asymmetry values obtained from the WSO synoptic maps. The period of erroneous data in 1996{\thinspace}--{\thinspace}2001.5 \citep{Virtanen2016,Virtanen2017} is depicted by a weak dashed line in the upper panel and as a gap in the lower panel. Out of $479$ (this number does not include weak-field asymmetries for the erroneous data period). For WSO weak-field asymmetries, $440$ ($92\%$) are statistically significant. Note also the greatly similar evolution of asymmetries in WSO and SOLIS/VSM over the overlapping time since 2003.\\

\section{Comparing the weak-field asymmetry values of the six data-sets}
\label{Comparing the weak-field asymmetry values of the six data-sets}

\begin{figure}[htpb]
	\centering
	\includegraphics[width=\columnwidth]{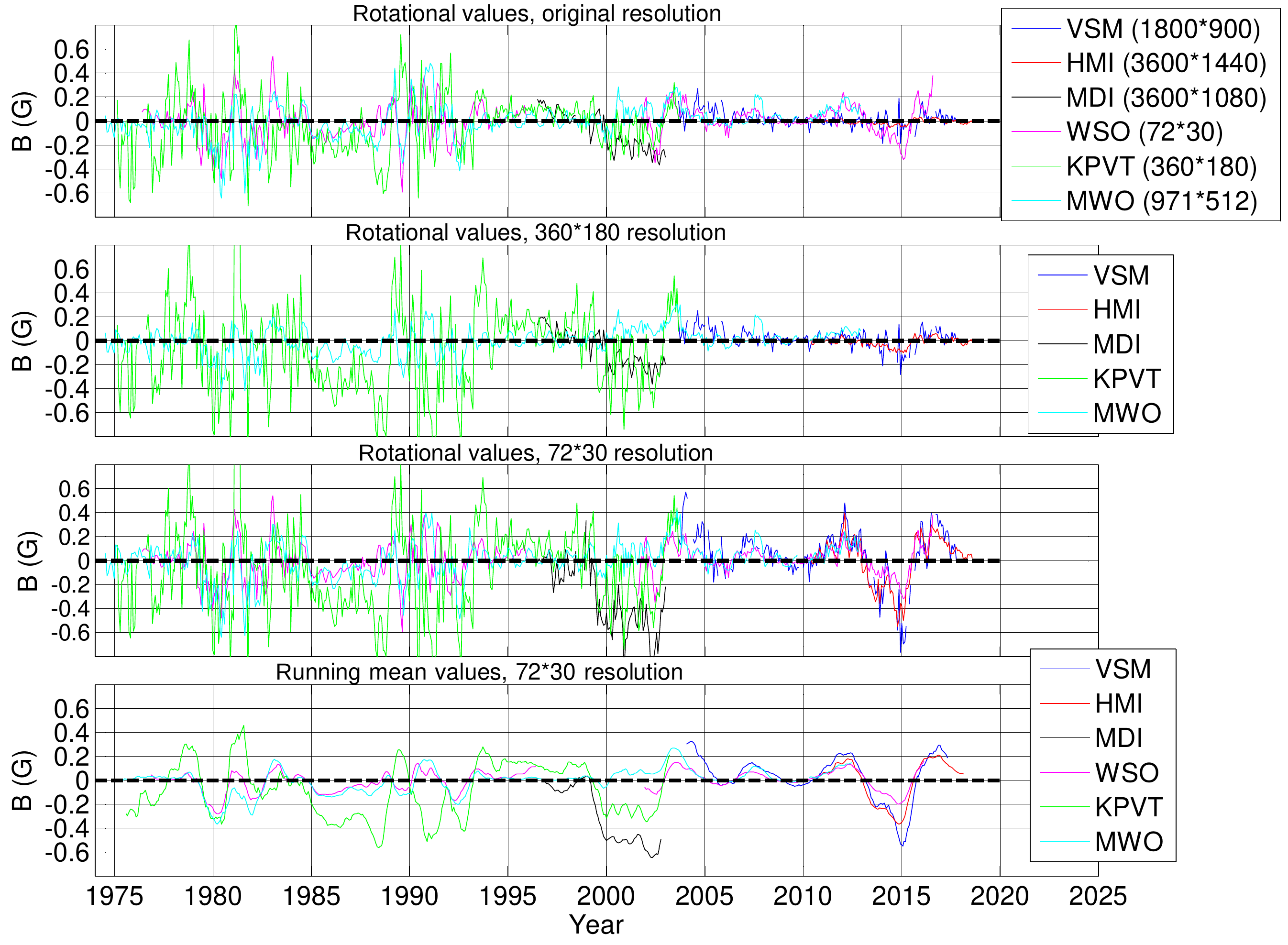}
	\caption{Weak-field asymmetries of the five data-sets at different resolutions. Upper panel gives rotational values of weak-field asymmetries obtained from VSM (blue line), HMI (red line), MDI (black line), WSO (magenta line), KPVT (green line) and MWO (cyan line) original resolution synoptic maps. The periods of erroneous WSO (from 1996{\thinspace}--{\thinspace}2001.5) and MDI weak-field asymmetry values  after 2003 are ignored.  Second panel gives rotational values of weak-field asymmetries for VSM (blue line), HMI (red line), MDI (black line), KPVT (green line) and MWO (cyan line) synoptic maps at 360*180 resolution. Third panel gives the rotational values of weak-field asymmetry of VSM, HMI, MDI, WSO, KPVT, and MWO synoptic maps at 72*30 resolution. Bottom panel gives the 13-rotation running mean values of third panel.}
	\label{fig:five_data_sets_zero_level_error}
\end{figure}

We now compare the weak-field asymmetries of all the six data-sets repeated in Figure \ref{fig:five_data_sets_zero_level_error}. The upper panel shows the rotational values of the weak-field asymmetries derived from the original maps of the six data-sets. Note that the original resolution of the different data-sets varies a lot. Second panel presents the rotational values of the weak-field asymmetries at the common 360*180 resolution for VSM, HMI, MDI, KPVT (for KPVT this is the original resolution) and MWO. Third panel depicts the weak-field asymmetry values at 72*30 resolution for all the six data-sets. (This is the original resolution of WSO). The bottom panel presents the 13-rotation running mean values of the 72*30 rotational values depicted in the third panel. (Running mean values are calculated with the condition that at least 7 out of 13 rotations must have a measured value for each running mean value.)\\

As can be seen from the third panel of Figure \ref{fig:five_data_sets_zero_level_error}, the weak-field asymmetry values derived from the six data-sets agree very well for the lowest-resolution 72*30 maps. The agreement is even better visible in the 13-rotation running mean values (bottom panel), where the annual variation seen in the third panel in some data-sets is averaged out. In fact, the weak-field asymmetries vary fairly similarly over the 44-year interval for all available data-sets shown in Figure \ref{fig:five_data_sets_zero_level_error}. The KPVT weak-field asymmetries show larger values than other data-sets at the same resolution, especially during 512-channel magnetograph period. MDI shows excessively large weak-field asymmetries since about 2000 which is most likely related to shutter noise \citep{Liu2004}. The values of weak-field asymmetries most often increase as the resolution of synoptic maps becomes lower, which is best seen for HMI and VSM.\\

We calculated correlation coefficients ($r$) and the corresponding p-values of the rotational weak-field asymmetries among the five data-sets at different resolutions. Table \ref{Corr_coef} shows the calculated $r$ (and p in parenthesis) between VSM (for three maps shown in Figure \ref{fig:solis_zero_level_error}) and HMI (for five types of maps shown in Figure \ref{fig:hmi_zero_level_error}), between VSM (three types) and MWO (for three types of maps shown in Figure \ref{fig:MWO_zero_level_error}), between VSM (three types) and WSO, between KPVT (for two types of maps shown in Figure \ref{fig:solis_zero_level_error}) and MWO (three types), between KPVT (two types) and WSO, between WSO and HMI (five types), and between WSO and MWO (three types). Note that when calculating $r$, the WSO weak-field asymmetries in 1996{\thinspace}--{\thinspace}2001.5 are ignored. Also we do not calculate $r$ between MDI and any other data-sets, as time interval of reliable MDI weak-field asymmetries is too short. \\

As can be seen from Table \ref{Corr_coef}, the correlation between VSM and HMI weak-field asymmetries is highest ($r=0.9$) for the low-resolution maps (VSM 72*30 vs HMI 120*48 and VSM 72*30 vs HMI 72*30). Overall, the weak-field asymmetries of the two lowest-resolution VSM synoptic maps correlate extremely significantly ($p<10^{-7}$) with the weak-field asymmetry values of all the five types of the HMI synoptic maps, and with WSO maps. The weak-field asymmetry values of all the five types of the HMI synoptic maps are even more significant ($p<10^{-14}$) with the WSO weak-field asymmetries and the significance mainly increases as the resolution of HMI synoptic map decreases. The significance of correlations between VSM and WSO also greatly varies with VSM resolution. The lowest-resolution VSM synoptic map yields also significant correlations ($p<10^{-4}$) with the weak-field asymmetry values of all the three types of the MWO synoptic maps. The same is true for the lowest-resolution KPVT synoptic maps with MWO ($p<10^{-4}$). The correlations between the two types of KPVT synoptic maps and WSO synoptic maps are also significant ($p<10^{-2}$). The WSO-MWO correlations are also extremely significant ($p<10^{-25}$).\\

\begin{table*}[h]
			\caption{Correlation coefficients between the weak-field asymmetry values from two simultaneous data-sets. Corresponding p-values are given in parenthesis.}
\resizebox{\columnwidth}{!}{%

	\centering
		\begin{tabular}{|l|l|l|l|l|l|l|l|l|l|l|}
			  \hline
vs& \shortstack{HMI \\ (3600*1440)}&\shortstack{HMI\\(360*180)}&\shortstack{HMI\\(180*72)}&\shortstack{HMI\\(120*48)}&\shortstack{HMI\\(72*30)}&\shortstack{MWO\\(971*512)}&\shortstack{MWO\\(360*180)}&\shortstack{MWO\\(72*30)}&\shortstack{WSO\\(72*30)}\\ 
\hline
VSM (1800*900)&\shortstack{0.41 \\($6.10^{-5}$)}&\shortstack{0.36 \\($5.10^{-4}$)}&\shortstack{0.26 \\(0.01)}&\shortstack{0.30 \\($4.10^{-3}$)}&\shortstack{0.30 \\($4.10^{-3}$)}&\shortstack{-0.13\\($0.19$)}&\shortstack{$-5.10^{-4}$\\(1.0)}&\shortstack{$-2.10^{-3}$\\(0.98)}&\shortstack{0.17\\($3.10^{-2}$)}\\ 
\hline
VSM (360*180)&\shortstack{ 0.58 \\($2.10^{-9}$)}&\shortstack{0.62\\($5.10^{-11}$)}&\shortstack{0.57\\($1.10^{-9}$)}&\shortstack{0.60\\($3.10^{-10}$)}&\shortstack{0.58\\($3.10^{-9}$)}&\shortstack{-0.04\\(0.7)}&\shortstack{-0.01\\(0.9)}&\shortstack{-0.02\\(0.89)}&\shortstack{0.43\\($4.10^{-8}$)}\\ 
\hline
VSM (72*30)&\shortstack{ 0.60 \\($2.10^{-10}$)} &\shortstack{0.77\\($6.10^{-19}$)}&\shortstack{0.88\\($1.10^{-27}$)} &\shortstack{0.90\\($4.10^{-34}$)} &\shortstack{0.90\\($2.10^{-33}$)}&\shortstack{0.47\\($3.10^{-7}$)}&\shortstack{0.38\\($4.10^{-5}$)}&\shortstack{0.5\\($2.10^{-8}$)}&\shortstack{0.82\\($6.10^{-39}$)}\\ 
\hline
KPVT (360*180)&-&-&-&-&-&\shortstack{0.16 \\($2.10^{-3}$)}&\shortstack{0.12 \\($2.10^{-1}$)}&\shortstack{0.14 \\($7.10^{-3}$)}&\shortstack{0.16 \\($5.10^{-3}$)}\\ 
\hline
KPVT (72*30)&-&-&-&-&-&\shortstack{0.20 \\($8.10^{-5})$}&\shortstack{0.23 \\($6.10^{-6}$)}&\shortstack{0.24 \\($1.10^{-6}$)}&\shortstack{0.37 \\($7.10^{-11}$)}\\
\hline
WSO (72*30)&\shortstack{ 0.74 \\($2.10^{-15}$)}&\shortstack{0.87 \\($3.10^{-27}$)}&\shortstack{0.90 \\($6.10^{-32}$)}&\shortstack{0.92\\($6.10^{-36}$)}&\shortstack{0.91\\($2.10^{-32}$)}&\shortstack{0.5\\
($6.10^{-31}$)}&\shortstack{0.5\\($1.10^{-26}$)}&\shortstack{0.57\\($2.10^{-36}$)}&\\ \cline{1-10}
\hline
		\end{tabular}

     }

\label{Corr_coef}

\end{table*}

\section{Discussion and conclusions}
\label{Discussion and conclusions}

In this paper we have studied the distribution of the weak values of the photospheric magnetic field using several data-sets. We calculated the possible shifts of the maximum of the Gaussian-fitted distribution of weak-field values from zero, here called the weak-field asymmetries and studied their statistical significance, temporal occurrence and similarity among the many data sets. \\

 We compared the weak-field asymmetries obtained from high- (original), medium- and low-resolution versions of a given synoptic map. We found that weak-field asymmetry values are mostly fairly small for high-resolution maps, but increase systematically with decreasing resolution. A large fraction (from 40\% to 95\%) of weak-field asymmetry values are statistically significant at any resolution. This percentage is large even for the highest-resolution maps and increases systematically for lower-resolution maps. Moreover, we have shown that the rotations with significant non-zero weak-field asymmetries already appear at the highest resolution, and have mostly the same sign for the different resolutions.  \\

We calculated the correlation coefficients and p-values for the weak-field asymmetries between all the different resolutions of the five data-sets (MDI was left out). The weak-field asymmetries vary fairly similarly over the 44-year interval in the different data-sets. Significance of correlations of weak-field asymmetries is very high for low-resolution maps. This good agreement between the many data-sets is outstanding, taking into account the many differences between the data-sets due, e.g., to instrumental, measurement and calibration differences, as well as differences in the construction of synoptic maps. This gives strong evidence for the physical, non-random nature of the weak-field asymmetry values.  \\

The common assumption that the maximum weak-field distribution should be at zero field is based on the idea that measurements are spatially too inaccurate to resolve the ultimate scale of magnetic field elements. A zero maximum of the field distribution means that either the magnetograph has, due to insufficient spectral resolution, a threshold below which magnetic field elements are undetectable and the region is observationally unmagnetized for the instrument \citep{Harvey1999}, or that the magnetograph has an insufficient spatial resolution so that the fluxes of opposite polarity fields within the spatial scale of that instrument cancel each other. \\

We compared the weak-field asymmetries obtained from high- and low-resolution synoptic maps and found that the weak-field asymmetries for high-resolution maps are typically considerably smaller than for low-resolution maps. Our results, especially for HMI and VSM weak-field asymmetries show that for reliable high-resolution maps, the weak-field distribution has its maximum at a rather small field value which, however, still can be statistically significant and typically has the same sign as the low-resolution maps. For low-resolution synoptic maps, the maximum of the distribution shows very often a considerably larger weak-field asymmetry value, which is highly significant, and has typically the same sign (and quite similar absolute values) for the different data-sets.\\

 We note that, although not explicitly shown in this paper, we have also studied the asymmetries of the individual magnetograms (from which the synoptic maps are constructed) and found out that, while the histogram peaks of these images are random, the Gaussian fits of the weak-field distribution yield closely similar (statistically significant), physical weak-field asymmetries as the synoptic maps. This further verifies the reliability of the results obtained from synoptic maps.\\


The appearance of large weak-field asymmetries for low-resolution synoptic maps indicates that we start seeing an asymmetric distribution of magnetic fields at the supergranulation scale. The average size of supergranulation diameter is about $2.4^{o}$ or 30 000 km \citep[see, e.g.,][]{Pevtsov2016,Rincon2018}. In the case of synoptic maps, this scale corresponds to the resolution of about 150*75. Accordingly, the effect of supergranulation in the weak-field distribution is not well detected in synoptic maps which have a resolution higher than 150*75.\\

Concluding, our results suggest that the (non-zero) weak-field asymmetries reflect a real feature in the distribution of positive and negative weak-field values produced most effectively at the supergranulation scale, which can be best seen in medium- and low-resolution synoptic maps.

\begin{acknowledgements}
{\emph{Acknowledgements.} We acknowledge the financial support by the Academy of Finland to the ReSoLVE Centre of Excellence. Wilcox Solar Observatory data used in this study was obtained via the web site  \url{http://wso.stanford.edu} courtesy of J.T. Hoeksema. This study includes data from the synoptic program at the 150-Foot Solar Tower of the Mt. Wilson Observatory, which is acknowledged. NSO/Kitt Peak magnetic data used here are produced cooperatively by NSF/NOAO, NASA/GSFC and NOAA/SEL. Data were acquired by SOLIS instruments operated by NISP/NSO/AURA/NSF. SOHO/MDI is a project of international cooperation between ESA and NASA. HMI data are courtesy of the Joint Science Operations Center (JSOC) Science Data Processing team at Stanford University.\\}

Data used in this study was obtained from the following web sites:\\
WSO: \url{http://wso.stanford.edu}\\
MWO: \url{ftp://howard.astro.ucla.edu/pub/obs/synoptic_charts}\\
Kitt Peak: \url{ftp://solis.nso.edu/kpvt/synoptic/mag/}\\
MDI: \url{http://soi.stanford.edu/magnetic/synoptic/carrot/M_Corr}\\
HMI: \url{http://jsoc.stanford.edu/data/hmi/synoptic}\\
SOLIS: \url{http://solis.nso.edu/0/vsm/vsm_maps.php}\\

\end{acknowledgements}


\begin{thebibliography}{}
	\expandafter\ifx\csname natexlab\endcsname\relax\def\natexlab#1{#1}\fi
	\providecommand{\url}[1]{\href{#1}{#1}}
	
	\bibitem[{{Babcock}(1953)}]{Babcock1953}
	{Babcock}, H.~W. 1953, \apj, 118, 387
	
	\bibitem[{Bertello {et~al.}(2014)Bertello, Pevtsov, Petrie, \&
		Keys}]{Bertello2014}
	Bertello, L., Pevtsov, A.~A., Petrie, G. J.~D., \& Keys, D. 2014, Solar
	Physics, 289, 2419
	
	\bibitem[{Getachew {et~al.}(2017)Getachew, Virtanen, \& Mursula}]{Getachew2017}
	Getachew, T., Virtanen, I., \& Mursula, K. 2017, Solar Physics, 292, 174.
	\newblock \url{https://doi.org/10.1007/s11207-017-1198-9}
	
	\bibitem[{{Harvey} \& {Munoz-Jaramillo}(2015)}]{Harvey2015}
	{Harvey}, J., \& {Munoz-Jaramillo}, A. 2015, in AAS/AGU Triennial Earth-Sun
	Summit, Vol.~1, AAS/AGU Triennial Earth-Sun Summit, 111.02
	
	\bibitem[{Harvey \& White(1999)}]{Harvey1999}
	Harvey, K.~L., \& White, O.~R. 1999, The Astrophysical Journal, 515, 812.
	\newblock \url{http://stacks.iop.org/0004-637X/515/i=2/a=812}
	
	\bibitem[{Hayashi {et~al.}(2015)Hayashi, Hoeksema, Liu, Bobra, Sun, \&
		Norton}]{Hayashi2015}
	Hayashi, K., Hoeksema, J.~T., Liu, Y., {et~al.} 2015, Solar Physics, 290, 1507.
	\newblock \url{https://doi.org/10.1007/s11207-015-0686-z}
	
	\bibitem[{{Hoeksema}(1984)}]{Hoeksema1984}
	{Hoeksema}, J.~T. 1984, PhD thesis, Stanford Univ., CA.
	
	\bibitem[{{Hoeksema}(2010)}]{Hoeksema2010}
	{Hoeksema}, J.~T. 2010, in IAU Symposium, Vol. 264, Solar and Stellar
	Variability: Impact on Earth and Planets, ed. A.~G. {Kosovichev}, A.~H.
	{Andrei}, \& J.-P. {Rozelot}, 222--228
	
	\bibitem[{Hoeksema {et~al.}(2014)Hoeksema, Liu, Hayashi, Sun, Schou, Couvidat,
		Norton, Bobra, Centeno, Leka, Barnes, \& Turmon}]{Hoeksema2014}
	Hoeksema, J.~T., Liu, Y., Hayashi, K., {et~al.} 2014, Solar Physics, 289, 3483
	
	\bibitem[{{Howard} {et~al.}(1983){Howard}, {Boyden}, {Bruning}, {Clark},
		{Crist}, \& {Labonte}}]{Howard1983}
	{Howard}, R., {Boyden}, J.~E., {Bruning}, D.~H., {et~al.} 1983, \solphys, 87,
	195
	
	\bibitem[{{Howard}(1989)}]{Howard1989}
	{Howard}, R.~F. 1989, \solphys, 123, 271
	
	\bibitem[{{Liu} {et~al.}(2004){Liu}, {Zhao}, \& {Hoeksema}}]{Liu2004}
	{Liu}, Y., {Zhao}, X., \& {Hoeksema}, J.~T. 2004, \solphys, 219, 39
	
	\bibitem[{{Liu} {et~al.}(2012){Liu}, {Hoeksema}, {Scherrer}, {Schou},
		{Couvidat}, {Bush}, {Duvall}, {Hayashi}, {Sun}, \& {Zhao}}]{Liu2012}
	{Liu}, Y., {Hoeksema}, J.~T., {Scherrer}, P.~H., {et~al.} 2012, \solphys, 279,
	295
	
	\bibitem[{{Livingston} {et~al.}(1976){Livingston}, {Harvey}, {Pierce},
		{Schrage}, {Gillespie}, {Simmons}, \& {Slaughter}}]{Livingston1976}
	{Livingston}, W.~C., {Harvey}, J., {Pierce}, A.~K., {et~al.} 1976, \ao, 15, 33
	
	\bibitem[{{Pevtsov} {et~al.}(2016){Pevtsov}, {Virtanen}, {Mursula}, {Tlatov},
		\& {Bertello}}]{Pevtsov2016}
	{Pevtsov}, A.~A., {Virtanen}, I., {Mursula}, K., {Tlatov}, A., \& {Bertello},
	L. 2016, \aap, 585, A40
	
	\bibitem[{{Riley} {et~al.}(2014){Riley}, {Ben-Nun}, {Linker}, {Mikic},
		{Svalgaard}, {Harvey}, {Bertello}, {Hoeksema}, {Liu}, \&
		{Ulrich}}]{Riley2014}
	{Riley}, P., {Ben-Nun}, M., {Linker}, J.~A., {et~al.} 2014, \solphys, 289, 769
	
	\bibitem[{Rincon \& Rieutord(2018)}]{Rincon2018}
	Rincon, F., \& Rieutord, M. 2018, Living Reviews in Solar Physics, 15, 6
	
	\bibitem[{{Scherrer} {et~al.}(1995){Scherrer}, {Bogart}, {Bush}, {Hoeksema},
		{Kosovichev}, {Schou}, {Rosenberg}, {Springer}, {Tarbell}, {Title},
		{Wolfson}, {Zayer}, \& {MDI Engineering Team}}]{Scherrer1995}
	{Scherrer}, P.~H., {Bogart}, R.~S., {Bush}, R.~I., {et~al.} 1995, \solphys,
	162, 129
	
	\bibitem[{Schou {et~al.}(2012)Schou, Scherrer, Bush, Wachter, Couvidat,
		Rabello-Soares, Bogart, Hoeksema, Liu, Duvall, Akin, Allard, Miles, Rairden,
		Shine, Tarbell, Title, Wolfson, Elmore, Norton, \& Tomczyk}]{Schou2012}
	Schou, J., Scherrer, P.~H., Bush, R.~I., {et~al.} 2012, Solar Physics, 275, 229
	
	\bibitem[{{Svalgaard} {et~al.}(1978){Svalgaard}, {Duvall}, \&
		{Scherrer}}]{Svalgaard1978}
	{Svalgaard}, L., {Duvall}, Jr., T.~L., \& {Scherrer}, P.~H. 1978, \solphys, 58,
	225
	
	\bibitem[{{Svalgaard} {et~al.}(2011){Svalgaard}, {Hannah}, \&
		{Hudson}}]{Svalgaard2011}
	{Svalgaard}, L., {Hannah}, I.~G., \& {Hudson}, H.~S. 2011, \apj, 733, 49
	
	\bibitem[{{Svalgaard} \& {Wilcox}(1976)}]{Svalgaard1976}
	{Svalgaard}, L., \& {Wilcox}, J.~M. 1976, \solphys, 49, 177
	
	\bibitem[{{Ulrich} {et~al.}(2002){Ulrich}, {Evans}, {Boyden}, \&
		{Webster}}]{Ulrich2002}
	{Ulrich}, R.~K., {Evans}, S., {Boyden}, J.~E., \& {Webster}, L. 2002, \apjs,
	139, 259
	
	\bibitem[{Ulrich \& Tran(2013)}]{Ulrich2013}
	Ulrich, R.~K., \& Tran, T. 2013, The Astrophysical Journal, 768, 189.
	\newblock \url{http://stacks.iop.org/0004-637X/768/i=2/a=189}
	
	\bibitem[{{Virtanen} \& {Mursula}(2016)}]{Virtanen2016}
	{Virtanen}, I., \& {Mursula}, K. 2016, \aap, 591, A78
	
	\bibitem[{{Virtanen} \& {Mursula}(2017)}]{Virtanen2017}
	---. 2017, \aap, 604, A7
	
	\bibitem[{{Wilcox} \& {Howard}(1968)}]{Wilcox1968}
	{Wilcox}, J.~M., \& {Howard}, R. 1968, \solphys, 5, 564
	
\end{thebibliography}

\end{document}